\UseRawInputEncoding
\documentclass[journal]{IEEEtran}

\usepackage{amsmath,amsfonts}
\usepackage{amssymb}
\usepackage{algorithmic}
\usepackage{algorithm}
\usepackage{array}
\usepackage[caption=false,font=normalsize,labelfont=sf,textfont=sf]{subfig}
\usepackage{textcomp}
\usepackage{stfloats}
\usepackage{url}
\usepackage{verbatim}
\usepackage{xcolor}
\usepackage{booktabs}
\usepackage{graphicx}
\usepackage{cite}
\usepackage{xcolor}
\usepackage{etoolbox}

\makeatletter
\newcommand{\bluebibitems}{}

\pretocmd{\@bibitem}{%
  \ifinlist{#1}{\bluebibitems}{\color{blue}}{\normalcolor}%
}{}{}

\pretocmd{\@lbibitem}{%
  \ifinlist{#2}{\bluebibitems}{\color{blue}}{\normalcolor}%
}{}{}

\AtEndEnvironment{thebibliography}{\normalcolor}
\makeatother
\begin{document}

\title{Joint Transceiver and Group Index Modulation Design for Wideband Integrated Sensing and Communications Under Beam Squint}

\author{Kaiwen~Zheng,~\IEEEmembership{Student Member,~IEEE,}
        Shijian~Gao,~\IEEEmembership{Member,~IEEE}
    \thanks{Kaiwen Zheng and Shijian Gao are with the Internet of Things Thrust, Information Hub, The Hong Kong University of Science and Technology (Guangzhou), Guangzhou, China (E-mail: kzheng315@connect.hkust-gz.edu.cn; shijiangao@hkust-gz.edu.cn).}
}
\maketitle

\begin{abstract}
The combination of ultra-wide bandwidth and large-scale multiple-input multiple-output arrays with frequency-independent phase shifters induces beam squint, causing frequency and distance dependent array gain variations across subcarriers and degrading integrated sensing and communications performance. Transceiver beamforming and group index modulation (GIM) can alleviate the impact of beam squint on communication and sensing performance. However, their designs are coupled in the frequency domain, as beamforming shapes the subcarrier responses on which GIM grouping and activation depend. To address this coupling, we propose a joint transceiver beamforming and GIM design that maximizes a weighted combination of normalized worst-case communication and sensing distances. An alternating optimization algorithm jointly refines the transceiver beamformers and GIM configuration. Simulation results show that the proposed joint design reduces bit error rate in the high signal-to-noise ratio region and achieves a more favorable communication–sensing trade-off, while maintaining an average sensing AUC comparable to that of the considered GIM based benchmark schemes.
\end{abstract}


\section{Introduction}
\label{sec:intro}
\IEEEPARstart{W}{ith} the evolution of 6G, integrated sensing and communications (ISAC) has emerged as a key paradigm for multi-Gbps transmission and centimeter-level resolution \cite{9830717}. Millimeter-wave/sub-terahertz bands and massive multiple-input multiple-output (MIMO) provide the bandwidth and spatial resolution, while their large apertures make mixed near- and far-field operation increasingly relevant \cite{NBYmaxmin2023,10886954}. In wideband systems, large arrays suffer from beam squint, which causes frequency-dependent angular deviations in the far field and range-angle focal shifts in the near field, thereby degrading communication gains and sensing focus \cite{GFFBeamSquint2023,10886954}.

Existing beam squint studies have primarily addressed frequency-dependent mismatch through wideband beamforming. Far field methods either suppress angular deviations to maintain array gains across subcarriers \cite{NBYmaxmin2023} or exploit them for sensing-oriented beam scanning \cite{LHLYOLO2024}, while hybrid field designs additionally account for the range-dependent focal shifts \cite{GFFBeamSquint2023,10886954}. By reshaping the spectrum response, these methods alleviate communication gain loss and sensing defocusing while retaining a prescribed full-subcarrier orthogonal frequency-division multiplexing (OFDM) structure. In parallel, OFDM with index modulation (OFDM-IM) activates only a subset of subcarriers and conveys additional information through their indices \cite{6587554}. Subsequent studies improve its reliability by assigning subcarriers according to their channel conditions, constructing activation codebooks with favorable distance properties, or adapting the index signaling to frequency-selective channels \cite{MaSubcarrier2016,DangLexicographic2018,ZhengMaASIM2026}. IM has also been extended to ISAC, where index-bearing waveform components provide additional communication dimensions while contributing to sensing observations \cite{ElbirIMISACReview2024,10559944}. Nevertheless, the joint dependence of transceiver beamforming and IM design under hybrid field beam squint remains insufficiently explored.

To address this, we propose a joint transceiver beamforming and grouped index modulation (GIM) design for wideband ISAC scenes. The beamforming-dependent subcarrier-gain profile links spectrum processing with frequency-domain index mapping. Beamforming alleviates beam squint and shapes the communication and sensing responses across subcarriers, whereas subcarrier grouping and activation set design in GIM determine how these responses are selected and combined. Accordingly, the transceiver beamformer and the GIM configuration are co-optimized by maximizing a weighted combination of normalized worst-case communication and sensing distances. The resulting mixed continuous-discrete problem is solved using a safeguarded alternating procedure that combines spectrum-variable updates with grouping and activation set search. 
A concise performance analysis relates the communication distance to the pairwise error probability and the sensing distance to the area under the curve (AUC) of the matched-log-likelihood ratio (LLR) detector. Simulation results demonstrate improved high signal-to-noise ratio (SNR) bit error rate (BER) and a favorable communication--sensing trade-off while maintaining average sensing performance comparable to the considered baselines.

\section{System Model}
This section presents the signal model of a wideband ISAC system employing GIM for simultaneous multiuser communication and target sensing under hybrid-field propagation.

\subsection{GIM transmission}
Consider a wideband MIMO-OFDM ISAC system. The base station (BS) employs separate $N_t$ and $N_r$ half-wavelength uniform linear arrays (ULAs) for transmission and reception, respectively. It simultaneously serves $K_c$ single-antenna communication users $k\in\mathcal K_c$ at $(r_k,\theta_k)$ and probes $K_s$ candidate sensing locations $u\in\mathcal K_s$ at $(r_u,\theta_u)$ in hybrid field. The system operates at central frequency $f_c$ with bandwidth $B$, partitioned into $M$ orthogonal subcarriers indexed by $m \in [-\frac{M}{2}, \frac{M}{2}-1]$. With analog phase shifters, each entry of the transmit precoder $\mathbf{W} = [\mathbf{w}_1, \dots, \mathbf{w}_{K_c}] \in \mathbb{C}^{N_t \times K_c}$ and the receive combiner $\mathbf{R} = [\mathbf{r}_1, \dots, \mathbf{r}_{K_s}] \in \mathbb{C}^{N_r \times K_s}$ is unit-modulus.

Based on this system. Let $\mathcal M$ denote the set of $M=JG$ subcarriers. They are partitioned into $J$ disjoint groups $\mathcal G_j=\{\pi_{j,1},\ldots,\pi_{j,G}\}$ satisfying $\bigcup_{j=1}^{J}\mathcal G_j=\mathcal M,\mathcal G_j\cap\mathcal G_{j'}=\varnothing,\quad j\neq j'.$ The grouping is represented by $\boldsymbol{\Pi}=[\pi_{j,g}]$. Each group employs an activation set $\mathcal Z_j=\{\mathbf z_{j,q}\}_{q=1}^{Q}$, where $\mathbf z_{j,q}\in\{0,1\}^{G}$ and $\|\mathbf z_{j,q}\|_0=A$. The $Q$ codewords are distinct and $Q\leq\binom GA$ is a power of two. For $m=\pi_{j,g}$, define $X_{j,q,m}=Z_{j,q,g}$, and otherwise $X_{j,q,m}=0$. Each active subcarrier uses the power factor $G/A$. The grouping $\boldsymbol{\Pi}$ and activation set collection $\mathbf Z\triangleq\{\mathcal Z_j\}_{j=1}^{J}$ are jointly optimized with the beamforming matrix $\mathbf W$ and the sensing matrix $\mathbf R$.

\subsection{Communication model}
Let $x_{k,m}$ denote the unit-power symbol transmitted for communication user $k$ on subcarrier $m$. Following \cite{10541333}, the hybrid field channel is modeled as
$[\mathbf h_{k,m}]_n=\frac{r_k}{r_{k,n}}e^{-j2\pi f_m(r_{k,n}-r_k)/c}$, where $c$ is the speed of light
and $r_{k,n}=\sqrt{r_k^2+\delta_n^2+2r_k\delta_n\sin\theta_k}$, $\delta_n=(n-\frac{N_t-1}{2})\frac{c}{2f_c}$, and $n \in[0,\ldots,N_t-1]$. For sufficiently large $r_k$, $r_{k,n}-r_k\approx\delta_n\sin\theta_k$ and $r_k/r_{k,n}\approx1$, recovering the far-field array response. Defining $g^c_{k,j,m}\triangleq\mathbf h_{k,m}^H\mathbf w_j$, the received signal is
\begin{equation}\label{eq:rx_signal}
y^c_{k,m}
=g^c_{k,k,m}x_{k,m}
+\sum_{\substack{j\neq k}}^{K_c}
g^c_{k,j,m}x_{j,m}
+n_{k,m},
\end{equation}
where $n_{k,m}\sim\mathcal{CN}(0,\sigma^2)$ is the additive white Gaussian noise.

\subsection{Sensing model}
Let $b_l\in{\{0,1\}}$ indicate whether a target is present at candidate sensing location $l$, where $b_l=1$ denotes target presence.
With $\mathbf x_m=[x_{1,m},\ldots,x_{K_c,m}]^T$, the combined echo associated with sensing location $u\in\mathcal K_s$ on subcarrier $m$ is given by\footnote{Without loss of generality, the reflection coefficient together with round-trip path loss is set as $1$}:
\begin{equation}\label{eq:sens_signal}
y^{\rm s}_{u,m}
=b_u{\mathbf g^s_{u,u,m}}^T\mathbf x_m
+\sum_{\substack{l\in\mathcal K_s\\l\neq u}}
b_l{\mathbf g^s_{u,l,m}}^T\mathbf x_m
+n^{\rm s}_{u,m},
\end{equation}
where $n^s_{u,m} \sim \mathcal{CN}(0, \sigma^2)$. The equivalent reflection channel is defined as
${\mathbf g^s_{u,l,m}}^T
\triangleq
\mathbf r_u^H\mathbf s_{l,m}
\mathbf h_{l,m}^H\mathbf W$,
where $\mathbf h_{l,m}$ follows the hybrid field response above evaluated at $(r_l,\theta_l)$, and
$\mathbf s_{l,m}\in\mathbb C^{N_r}$ is its receive-array counterpart obtained from the same model using $N_r$ receive elements.

\section{Problem Formulation}
This section develops communication and sensing performance metrics based on the signal models above and formulates
the joint beamforming and GIM design problem.

\subsection{Communication Metric}
Based on \eqref{eq:rx_signal}, the effective signal-to-interference-plus-noise ratio (SINR) of user $k$ on subcarrier $m$ is
\begin{equation}
\gamma_{c,k,m}=\frac{|g^c_{k,k,m}|^2}{\sum_{i\neq k}^{K_c}|g^c_{k,i,m}|^2+\sigma^2}.
\end{equation}
In each group, the receiver recovers both the activation index and the modulation symbols using a LLR detector \cite{6587554}. The reliability of this detection depends on the separation between competing received-signal hypotheses. Treating multiuser interference as Gaussian noise, the minimum SINR-weighted distance for user $k$ in group $j$ is defined as 
$D^c_{k,j}=\min_{\substack{\mathbf a,\widehat{\mathbf a}\in\mathcal C_j\\\mathbf a\neq\widehat{\mathbf a}}}\sum_{m\in\mathcal G_j}\gamma_{c,k,m}|a_m-\widehat a_m|^2,$
where $\mathcal C_j$ is the set of feasible transmitted symbol vectors in group $j$, with entries $a_m=X_{j,q,m}x_m$ for an activation codeword $q\in\{1,\ldots,Q\}$ and modulation symbols drawn from the adopted constellation. Thus, $D^c_{k,j}$ accounts for errors in both activation-index and symbol detection.
To protect the least favorable user and group, the system communication metric is defined as
\begin{equation}\label{eq:system_comm_distance}
D_c=\min_{\substack{k\in\mathcal K_c\\j\in\{1,\ldots,J\}}}D^c_{k,j}.
\end{equation}
At high SNR, errors are primarily associated with the closest competing hypotheses \cite{6587554}, so maximizing $D_c$ is expected to improve BER performance.

\subsection{Sensing Metric}
Since the BS knows the transmitted activation pattern and modulation symbols, target presence is detected using a waveform-matched LLR receiver.
For sensing location $u$, define the known echo template as $t_{u,m}=({\mathbf g^s_{u,u,m}})^T\mathbf x_m$. Approximating the aggregate interference and noise as $v_{u,m}\sim\mathcal{CN}(0,\nu_{u,m})$, independent across subcarriers, the detection hypotheses are $H_{0,u}:y^s_{u,m}=v_{u,m} $ and $H_{1,u}:y^s_{u,m}=t_{u,m}+v_{u,m}.$ Let $\mathcal I$ denote the active-subcarrier set. The LLR accumulated over $\mathcal I$ is
\begin{equation}\label{eq:sensing_llr}
\Lambda_u=2\operatorname{Re}\left\{\sum_{m\in\mathcal I}\frac{t_{u,m}^{*}y^s_{u,m}}{\nu_{u,m}}\right\}-d_u,
\end{equation}
where $d_u=\sum_{m\in\mathcal I}|t_{u,m}|^2/\nu_{u,m}$ is the instantaneous detection distance. For a fixed transmitted waveform, $\Lambda_u|\mathcal H_{0,u}\sim\mathcal N(-d_u,2d_u)$ and $\Lambda_u|\mathcal H_{1,u}\sim\mathcal N(d_u,2d_u)$.
The corresponding $\operatorname{AUC}_u=\Phi(\sqrt{d_u})$, where $\Phi(\cdot)$ is the standard Gaussian cumulative
distribution function. Thus, a larger detection distance yields a higher AUC under the assumed Gaussian model. The sensing SINR is defined as
\begin{equation}
\gamma_{s,u,m}=\frac{\|\mathbf g^s_{u,u,m}\|^2}{\sum_{\substack{l\in\mathcal K_s\\l\neq u}}b_l\|\mathbf g^s_{u,l,m}\|^2+\sigma^2},
\end{equation}
Accordingly, the sensing distance associated with activation codeword $q$ in group $j$ is defined as
$D^s_{u,j}(q)=\sum_{m\in\mathcal G_j}X_{j,q,m}\gamma_{s,u,m}.$
The system sensing metric protects the least favorable location and activation-codeword combination:
\begin{equation}\label{eq:system_sensing_distance}
D_s=\min_{u\in\mathcal K_s}\sum_{j=1}^{J}\min_{q\in\{1,\ldots,Q\}}D^s_{u,j}(q).
\end{equation}
Protecting the worst-case distance mitigates poor detection performance under unfavorable activation patterns, while its impact on average AUC may remain limited.
\subsection{Joint Design Problem Formulation}
The communication and sensing metrics depend jointly on the $\mathbf W$ and $\mathbf R$, the subcarrier
grouping $\boldsymbol{\Pi}$, and the activation set $\mathbf Z$.
To account for their different scales, the two metrics are normalized as $\eta_c=D_c/\bar D_c$ and $\eta_s=D_s/\bar D_s$,
where $\bar D_c$ and $\bar D_s$ are fixed reference distances obtained from the corresponding ideal single-link systems without beam squint.
The joint design problem is formulated as
\begin{equation}\label{prob:joint_design}
\tag{P0}
\begin{aligned}
\underset{\mathbf W,\mathbf R,\boldsymbol{\Pi},\mathbf Z}
{\operatorname{maximize}}\quad
&\lambda\eta_c+(1-\lambda)\eta_s\\
\operatorname{subject\ to}\quad
&(\mathbf W,\mathbf R)\in\mathcal A,\\
&(\boldsymbol{\Pi},\mathbf Z)\in\mathcal G.
\end{aligned}
\end{equation}
where $\lambda\in[0,1]$ controls the communication--sensing trade-off.
The set $\mathcal A$ imposes the constant-modulus constraints on the transmit and receive beamformers, while $\mathcal G$
contains the disjoint subcarrier partitions and distinct activation set specified in Section~II-A.

\section{Proposed Solution}
The proposed problem couples continuous beamforming variables with discrete grouping and activation set in GIM through the communication and sensing distances. An alternating optimization (AO) procedure is adopted: the beamformers are updated for fixed GIM configuration, followed by discrete refinement for fixed beamformers. Each candidate block update is accepted only if it improves the original objective $F_\lambda=\lambda\eta_c+(1-\lambda)\eta_s$.

\subsection{Beamforming Optimization}
For fixed $\boldsymbol{\Pi}$ and $\mathbf Z$, the feasible transmission hypotheses and activation patterns are fixed.
The remaining beamforming subproblem is nonsmooth because of the minimum operators in $D_c$ and $D_s$.
Following \cite{xu2001smoothing}, for a finite collection $\{p_i\}_{i=1}^{L}$, define
$\operatorname{smin}_{\mu}\{p_i\}_{i=1}^{L} =-\frac{1}{\mu}\log\sum_{i=1}^{L}e^{-\mu p_i},$
where $\mu>0$ controls the smoothing accuracy. The approximation becomes tight as $\mu\rightarrow\infty$. Replacing each minimum operator in $D_c$ and $D_s$ by $\operatorname{smin}_{\mu}$ gives their smooth approximations
$D_{c,\mu}$ and $D_{s,\mu}$, yielding
$\mathcal J_\mu(\mathbf W,\mathbf R) = \lambda\frac{D_{c,\mu}}{\bar D_c} +(1-\lambda)\frac{D_{s,\mu}}{\bar D_s},$
where the dependence on $\mathbf W$ and $\mathbf R$ is implicit through the subcarrier SINRs in the distance. Although smoothing removes the nondifferentiability, the beamforming variables remain coupled in the numerators and denominators of the SINRs, making direct optimization difficult. The quadratic transform of fractional programming (FP) \cite{SKMFP2018} is therefore employed to obtain a tractable equivalent formulation with auxiliary variables. Applying this transform to the communication and sensing SINRs introduces auxiliary variables $\boldsymbol\Omega=\{\boldsymbol\alpha,\boldsymbol\Xi\}$.
For fixed $\mathbf W$ and $\mathbf R$, their optimal updates are
\begin{equation}\label{eq:auxiliary_update}
\begin{aligned}
\alpha_{k,m}
&=\frac{g^c_{k,k,m}}{\sum_{i\neq k}|g^c_{k,i,m}|^2+\sigma^2},\\\boldsymbol\Xi_{u,m}&=\frac{\mathbf g^s_{u,u,m}}{\sum_{\substack{l\in\mathcal K_s\\l\neq u}}b_l\|\mathbf g^s_{u,l,m}\|^2+\sigma^2}.
\end{aligned}
\end{equation}
Substituting the transformed SINRs into $\mathcal J_\mu$ gives $\widetilde{\mathcal J}_\mu(\mathbf W,\mathbf R,\boldsymbol\Omega)$. For fixed $\boldsymbol\Omega$, this objective is differentiable and blockwise concave, while the constant-modulus constraints remain nonconvex. Projected block coordinate ascent (PBCA) therefore alternately updates $\mathbf W$ and $\mathbf R$ through gradient ascent followed by constant-modulus projection, where$[\mathcal P_{\mathbb C}(\mathbf D)]_{i,j}=[\mathbf D]_{i,j}/|[\mathbf D]_{i,j}|$.

\subsection{GIM Optimization}
For fixed $\mathbf W$ and $\mathbf R$, the subcarrier SINRs are fixed. The GIM configuration are initialized using
the normalized joint subcarrier score $w_m=\lambda\frac{\min_k\gamma_{c,k,m}}{\sum_{\ell\in\mathcal M}\min_k\gamma_{c,k,\ell}}+(1-\lambda)\frac{\min_u\gamma_{s,u,m}}{\sum_{\ell\in\mathcal M}\min_u\gamma_{s,u,\ell}}.$ Subcarriers are sorted in descending order of $w_m$ and successively assigned to the non-full group with the smallest accumulated score. Each group is initialized with a feasible max--min-Hamming-distance.
The activation set are then refined group by group by evaluating all feasible selections of $Q$ distinct constant-weight
patterns and retaining the selection that maximizes $F_\lambda$, with the other groups fixed. Starting from this initialization, random cross-group swaps are performed to refine the grouping.
At each trial, two subcarriers from different groups are randomly selected and exchanged, and only the activation set
of the two affected groups are refined using the same procedure. The resulting grouping and activation set are accepted if they increase $F_\lambda$; otherwise, the current solution is retained. The search terminates when a prescribed number of consecutive trials yields no improvement or the trial budget is exhausted.

\subsection{Convergence and Optimality Analysis}
The overall procedure is summarized in Algorithm~\ref{alg:joint_ao}. Since the outer AO iterations retain the objective of problem~\eqref{prob:joint_design}, their convergence can be established through the monotonicity and boundedness of $F_\lambda$. For fixed channels, the unit-modulus constraints and positive noise variance ensure finite uniform SINR bounds $\gamma_{c,k,m}\leq\Gamma_c$ and $\gamma_{s,u,m}\leq\Gamma_s$ over all feasible beamformers. Each communication distance contains $G$ subcarrier contributions, and each sensing activation pattern contains $JA$ active subcarriers. Therefore,
\[
F_\lambda\leq\frac{\lambda Gd_{\max}^2\Gamma_c}{\bar D_c}+\frac{(1-\lambda)JA\Gamma_s}{\bar D_s}\triangleq U_\lambda<\infty,
\]
where $d_{\max}^2$ is the maximum squared distance between elements of the modulation constellation augmented with zero. Consequently, feasible AO updates that do not decrease the original objective generate a convergent objective sequence. The following analysis examines this requirement for the two subproblem updates.

For fixed GIM configuration and $\mu>0$, the softmin approximation satisfies $0\leq F_\lambda-\mathcal J_\mu\leq\Delta_\mu =\frac{\lambda\log N_c}{\mu\bar D_c}+\frac{(1-\lambda)(\log K_s+J\log Q)}{\mu\bar D_s},$
where $N_c$ is the total number of communication distance terms included in the softmin approximation. Thus, increasing $\mu$ reduces the approximation error.
Let $\mathbf B=(\mathbf W,\mathbf R)$. The quadratic transform gives the equivalent representation such that $\mathcal J_\mu(\mathbf B)=\max_{\boldsymbol\Omega}\widetilde{\mathcal J}_\mu(\mathbf B,\boldsymbol\Omega).$
With exact auxiliary updates and sequential projected beamformer updates using step sizes that ensure ascent, the inner iterations satisfy
$\mathcal J_\mu(\mathbf B^{i+1})
\geq
\widetilde{\mathcal J}_\mu
(\mathbf B^{i+1},\boldsymbol\Omega^{i+1})
\geq
\widetilde{\mathcal J}_\mu
(\mathbf W^{i+1},\mathbf R^i,
\boldsymbol\Omega^{i+1})
\geq
\widetilde{\mathcal J}_\mu
(\mathbf B^i,\boldsymbol\Omega^{i+1})
=
\mathcal J_\mu(\mathbf B^i).$
Since $\mathcal J_\mu\leq F_\lambda\leq U_\lambda$, the nondecreasing inner objective sequence converges. To ensure monotonicity of the original objective, the projected candidate replaces the current beamformers only if it does not decrease $F_\lambda$.

For fixed beamformers, the GIM refinement preserves the grouping and activation constraints, retains the current configuration, and accepts only changes that strictly increase $F_\lambda$. Since the discrete feasible set is finite, only finitely many improving updates can be accepted, while the prescribed search budget ensures termination. Together with the beamforming acceptance rule, this yields $F_\lambda^{(t)}\leq F_\lambda^{(t+1)}\leq U_\lambda.$
Hence, the overall algorithm maintains feasibility and generates a nondecreasing, bounded sequence of original objective values, which therefore converges. Global optimality is not guaranteed because the algorithm uses blockwise updates and inexact subproblem solutions.

The complexity is $\mathcal O(C_{\rm init}+T_o[T_bC_{\rm BF}+T_gC_{\rm GIM}])$, where $C_{\rm init}$ is the initialization cost, $T_o$ counts outer iterations, and $T_b,T_g$ and $C_{\rm BF},C_{\rm GIM}$ denote the respective counts and costs of beamforming and groupwise GIM updates per outer iteration. Each GIM update evaluates $\binom{\binom{G}{A}}{Q}$ candidate sets.

\begin{algorithm}[t]
\caption{AO Algorithm for Joint Beamforming and GIM Design}
\label{alg:joint_ao}
\begin{algorithmic}[1]
\REQUIRE Smoothing parameter $\mu$, step sizes $\delta_w$, $\delta_r$, terminat-
ing ratio $\epsilon$, maximum iteration number $T_{\max}$, and a feasible initialization
$\{\mathbf W,\mathbf R,\boldsymbol{\Pi},\mathbf Z\}$.
\STATE Set $t=0$ and evaluate $F_\lambda$.
\REPEAT
    \STATE Store the current $\mathbf W$ and $\mathbf R$; set $i=0$.
    \REPEAT
        \STATE Update $\boldsymbol{\Omega}^{(i+1)}$ via \eqref{eq:auxiliary_update}.
        \STATE  Compute gradient {\footnotesize $\mathbf{G}_w^{(i)} \leftarrow \nabla_{\mathbf{W}} \tilde{\mathcal{J}}_{\mu}(\mathbf{W}^{(i)}, \mathbf{R}^{(i)}, \boldsymbol{\alpha}^{(i+1)})$}.
        \STATE $\mathbf W^{(i+1)}\leftarrow
        \frac{1}{\sqrt{N_t}}\mathcal P_{\mathbb C}
        (\mathbf W^{(i)}+\delta_w\mathbf G_w^{(i)})$.
        \STATE Compute gradient {\footnotesize $\mathbf{G}_r^{(i)} \leftarrow \nabla_{\mathbf{R}} \tilde{\mathcal{J}}_{\mu}(\mathbf{W}^{(i)}, \mathbf{R}^{(i)}, \boldsymbol{\Xi}^{(i+1)})$}
        \STATE $\mathbf R^{(i+1)}\leftarrow
        \frac{1}{\sqrt{N_r}}\mathcal P_{\mathbb C}
        (\mathbf R^{(i)}+\delta_r\mathbf G_r^{(i)})$.
        \STATE $i\leftarrow i+1$.
    \UNTIL{ $\frac{|\mathcal{J}_{\mu}^{(i)} - \mathcal{J}_{\mu}^{(i-1)}|}{|\mathcal{J}_{\mu}^{(i-1)}|} < \epsilon$ \textbf{or} $i \ge T_{\max}$ }
    \STATE Restore optimized $\mathbf W^*$ and $\mathbf R^*$.
    
    \STATE Set $\ell=0$, init $(\boldsymbol{\Pi}^{(0)},\mathbf Z^{(0)})$.
    \REPEAT
        \STATE Refine $\mathbf Z$ group by group for fixed $\boldsymbol{\Pi}$.
        \STATE Generate $\mathcal N$ by cross-group subcarrier swaps.
        \STATE {\footnotesize $(\boldsymbol{\Pi}^*,\mathbf Z^*)\leftarrow
        \arg\max_{(\widetilde{\boldsymbol{\Pi}},\widetilde{\mathbf Z})
        \in\mathcal N\cup\{(\boldsymbol{\Pi},\mathbf Z)\}}
        F_\lambda(\mathbf W,\mathbf R,
        \widetilde{\boldsymbol{\Pi}},\widetilde{\mathbf Z})$.}
        \STATE $\ell\leftarrow\ell+1$.
    \UNTIL{$(\boldsymbol{\Pi}^*,\mathbf Z^*)$ remains unchanged or $\ell \ge T_{max}$}
    
    \STATE Evaluate $F_\lambda^{(t+1)}$ and set $t\leftarrow t+1$.
\UNTIL{$|F_\lambda^{(t)}-F_\lambda^{(t-1)}|\leq\epsilon$
or $t\geq T_{\max}$}
\ENSURE $\mathbf W^{*},\mathbf R^{*},
\boldsymbol{\Pi}^{*},\mathbf Z^{*}$.
\end{algorithmic}
\end{algorithm}

\section{Simulation Results}
Default parameters are $f_c=0.1$ THz, $B=30$ GHz, $M=64$, $N_t=64$, $N_r=32$, $K_c=2$, $K_s=1$, and $(G,A,Q,J)=(4,2,4,16)$. The communication users are located at $(20~\mathrm{m},-45^\circ)$ and $(25~\mathrm{m},-20^\circ)$, while the sensing target is located at $(3~\mathrm{m},45^\circ)$. Binary phase-shift keying (BPSK) is adopted. The proposed joint design is compared with OFDM (full-subcarrier), OFDM-IM (contiguous-group), MO-IM-OFDM (interleaved multi-observation\cite{10438844}), LC-GIM (lexicographic-codebook\cite{DangLexicographic2018}), and S-IM-OFDM (sequence-superposed\cite{10559944}), together with the sequential GIM (without outer AO loop), comm-only \cite{NBYmaxmin2023} $(\lambda=1)$ and sens-only \cite{LYXIOTJ2025} $(\lambda=0)$ ablations of the proposed algorithm. The SNR is defined as $E_b/N_0$, where $E_b$ is the energy per information bit and $N_0$ is the noise power spectral density. 

\subsection{BER Performance Versus SNR}
\begin{figure*}[t]
  \centering
  \subfloat[]{%
    \includegraphics[
      width=0.45\textwidth,
      trim={0 1mm 0 1mm},
      clip
    ]{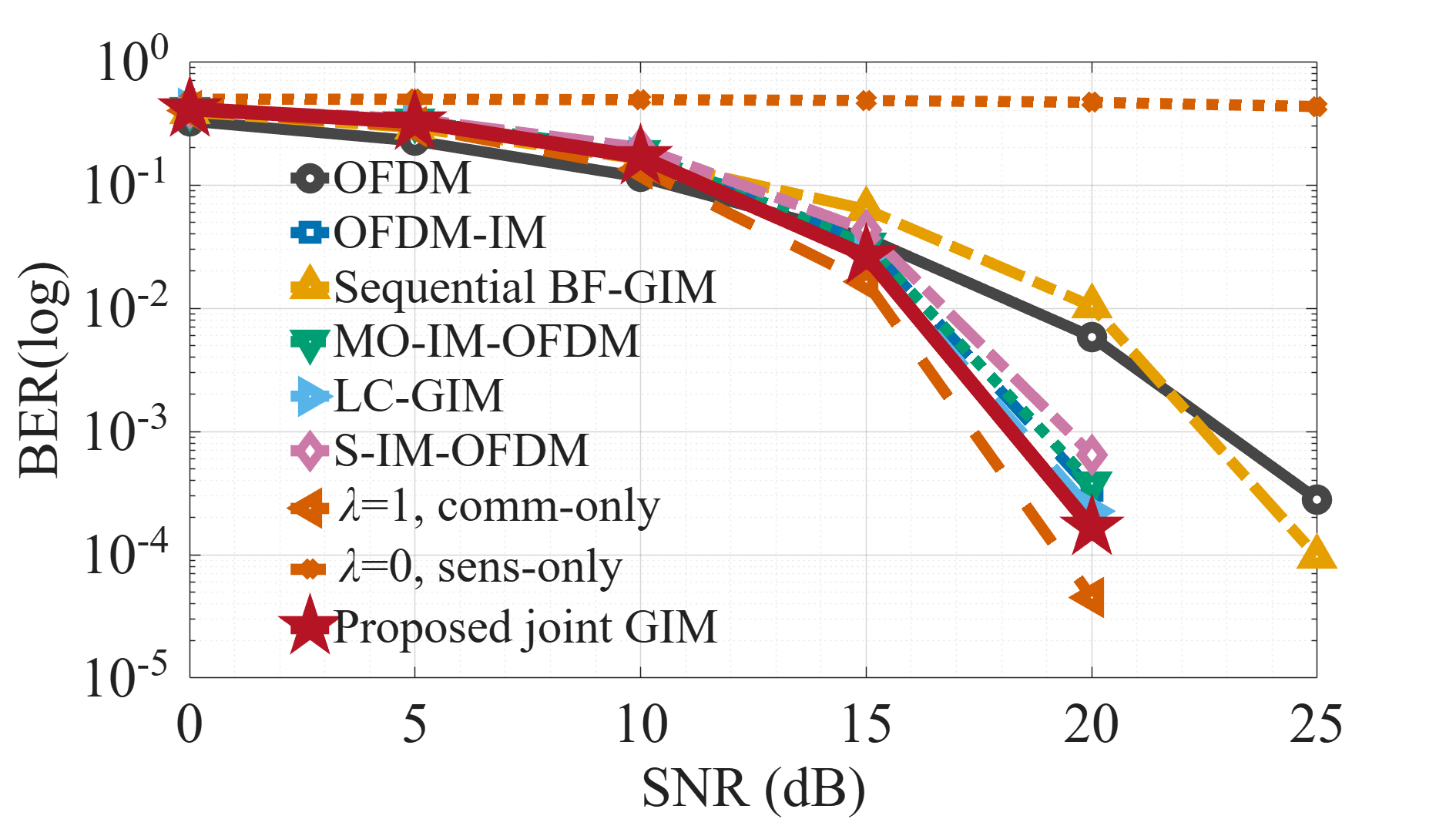}%
    \label{fig:comm_ber}%
  }\hfill
  \subfloat[]{%
    \includegraphics[
      width=0.45\textwidth,
      trim={0 1mm 0 1mm},
      clip
    ]{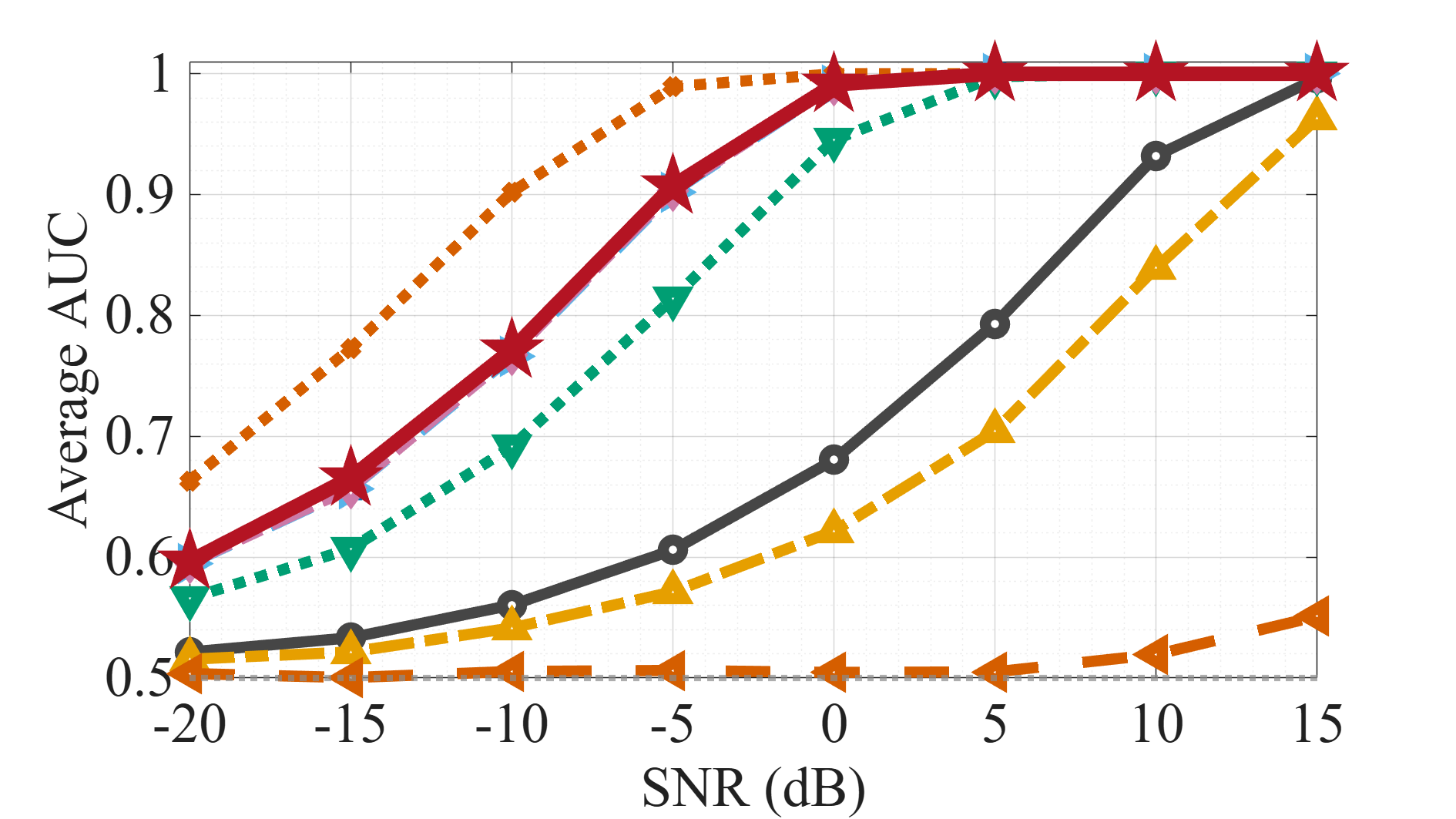}%
    \label{fig:sens_roc_auc}%
  }
  \caption{ISAC performance evaluation:
  (a) BER versus SNR;
  (b) AUC versus SNR.}
  \label{fig:overall_performance}
\end{figure*}

Fig.~\ref{fig:comm_ber} reports the multi-user average BER under BPSK signaling. At a BER of \(10^{-3}\), the proposed joint GIM provides approximately \(4.7\) dB and \(1\) dB gains over OFDM and OFDM-IM, respectively. Compared with its ablations, the joint design incurs about \(1.2\) dB loss relative to the communication-only design while substantially outperforming the sensing-only design, confirming the intended trade-off; it also gains about \(4\) dB over sequential joint GIM, demonstrating the benefit of the AO optimization framework and proving that the spectrum design is coupled with the GIM design. Moreover, it provides approximately \(0.2\!-\!1.5\) dB gains over MO-IM-OFDM, LC-GIM, and S-IM-OFDM, whose fixed interleaving, lexicographic codebook, or sensing-sequence allocation cannot jointly adapt the spectrum response and index mapping in the scenario.

\subsection{Sensing Performance Versus SNR}
Fig.~\ref{fig:sens_roc_auc} reports the average sensing AUC. The proposed joint GIM performs better than OFDM, while remaining comparable to OFDM-IM, LC-GIM, and S-IM-OFDM especially in moderate SNR. Among the proposed variants, the sensing-only design performs best, whereas the communication-only design remains close to random detection; the joint design recovers nearly all sensing capability while retaining its communication advantage, thereby confirming the intended trade-off. It also substantially outperforms sequential joint GIM and MO-IM-OFDM because the former lacks spectrum-GIM feedback and the latter distributes its sensing energy over multiple observations. Unlike high-SNR BER, which is dominated by minimum-distance error events, the sensing AUC is averaged over transmitted codewords; hence, the rarely occurring worst-codeword combination has limited influence on the average AUC, and its protection mainly improves robustness while leaving the average performance comparable to the baselines.

\subsection{Communication-Sensing Tradeoff Under Fixed SNR}
\begin{figure}[h]
  \centering
  \setlength{\abovecaptionskip}{2pt}
  \setlength{\belowcaptionskip}{-2pt}
  \includegraphics[width=0.8\columnwidth]{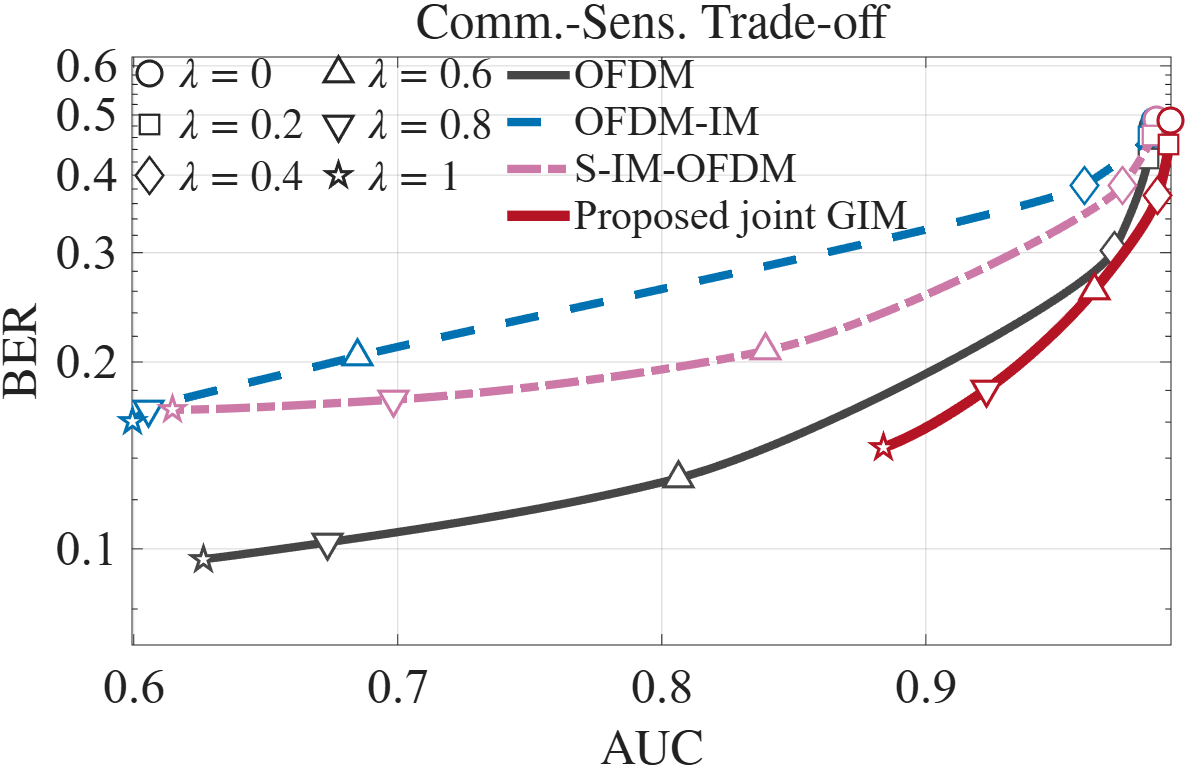}
  \caption{Communication-sensing trade-off at $\mathrm{SNR}=10$ dB.}
  \label{fig:trade-off}
\end{figure}

Fig.~\ref{fig:trade-off} illustrates the communication--sensing trade-off at a moderate SNR ($\rho=10$dB). OFDM-IM and S-IM-OFDM perform worse than OFDM in this regime because their index-detection gains have not yet compensated for the errors caused by sparse activation. In contrast, the proposed joint GIM jointly optimizes the beamformers and GIM design to better exploit the frequency-domain channel variations, thereby achieving a more favorable BER--AUC trade-off than OFDM. Together with the high-SNR BER results in Fig.~\ref{fig:comm_ber}, this observation suggests that the trade-off advantage will become more pronounced as the SNR increases.

\section{Conclusions}
This work investigated joint beamforming and GIM design for wideband MIMO-ISAC systems under hybrid-field beam squint. A GIM framework was developed to jointly optimize the transceiver beamformers and GIM according to normalized worst-case communication and sensing distances. An AO algorithm was further designed to coordinate the continuous beamforming variables and discrete GIM configuration while ensuring monotonic objective improvement. Simulation results showed that the proposed design improves high-SNR BER and achieves a more favorable communication–sensing trade-off than the considered OFDM and IM based benchmark schemes. Meanwhile, it maintains comparable average sensing AUC and improves the least favorable sensing performance. These results demonstrate that jointly adapting beamforming and frequency-domain index mapping can enhance communication reliability and sensing robustness without compromising overall sensing capability.

\bibliographystyle{IEEEtran}
\bibliography{IEEEabrv,ref} 
\end{document}